\PassOptionsToPackage{hyphens}{url} 
\documentclass[%
 aip,
 amsmath,amssymb,
 reprint,%
]{revtex4-2}
\usepackage{dcolumn,mathptmx,setspace}
\usepackage{bbm}
\usepackage[colorinlistoftodos]{todonotes}
\usepackage[utf8]{inputenc} 
\usepackage[T1]{fontenc}    

\usepackage[utf8]{inputenc} 
\usepackage[noend]{algpseudocode}

\usepackage{amssymb,amsmath,mathrsfs,stmaryrd,amsthm,mathtools,amsfonts,graphicx,hyperref,url,booktabs,nicefrac,comment,microtype,tikz-cd,lingmacros,nccmath,tree-dvips, bbm, bm, etoolbox,algorithm,caption,subcaption,color,enumitem,relsize,soul}

\usepackage{amsmath}

\begin{document}

\preprint{AIP/123-QED} 
\title[Nonlinear shifts and dislocations in financial market structure and composition]{Nonlinear shifts and dislocations in financial market structure and composition}

\author{Nick James}
\email{nick.james@unimelb.edu.au}
\affiliation{ 
School of Mathematics and Statistics, University of Melbourne, Victoria, 3010, Australia}%
\affiliation{Melbourne Centre for Data Science, University of Melbourne, Victoria, 3010, Australia}
\author{Max Menzies}
\email{max.menzies@alumni.harvard.edu}
\affiliation{%
Beijing Institute of Mathematical Sciences and Applications, Beijing, 101408, China}%

\date{February 26 2024}
\begin{abstract}

This paper develops new mathematical techniques to identify temporal shifts among a collection of US equities partitioned into a new and more detailed set of market sectors. Although conceptually related, our three analyses reveal distinct insights about financial markets, with meaningful implications for investment managers. First, we explore a variety of methods to identify nonlinear shifts in market sector structure and describe the mathematical connection between the measure used and the captured phenomena. Second, we study network structure with respect to our new market sectors and identify meaningfully connected sector-to-sector mappings. Finally, we conduct a series of sampling experiments over different sample spaces and contrast the distribution of Sharpe ratios produced by long-only, long-short and short-only investment portfolios. In addition, we examine the sector composition of the top-performing portfolios for each of these portfolio styles. In practice, the methods proposed in this paper could be used to identify regime shifts, optimally structured portfolios, and better communities of equities.

\end{abstract}

\maketitle

\begin{quotation}

Equity markets are highly nonstationary and continually change their behaviors over time. In recent years, many securities and asset classes have exhibited huge gains and/or fluctuations in their returns. In particular, equities exposed to the broad macroeconomic themes of artificial intelligence, decarbonization, and energy prices have benefited from growing investor interest. As investors turn to these market themes, however, other sectors are being overlooked and are experiencing capital outflows. These shifts in market dynamics can yield dislocations and create opportunities for arbitrage with respect to certain sectors. All these reasons make the time-varying study of stock prices more essential than ever. In addition, changes to the 21st century landscape of businesses require more nuanced sectors to divide the marketplace than traditional options such as the GICS classification system. In this work, we develop new methodologies to most appropriately identify nonlinear shifts in market behaviors, to reveal suitable market themes for portfolio diversification, and to optimally construct portfolios to perform best over long time horizons, all via a modern and sophisticated sector division.

\end{quotation}

\section{Introduction}

Modern financial portfolio management is an increasing complex endeavor. Sophisticated investment firms have numerous teams focused on strategic, dynamic and well-diversified asset selection over a variety of time frames, from periods of nanoseconds to years. In this paper, we look holistically over the long term, and address three aspects that remain consistently pertinent to portfolio selection: shifts in market structure and behavior over time, representation of various sectors in strong-performing portfolios over the long-term, and a framework for identifying communities of equities from a universe of potential candidates.

A significant body of literature focuses on the evolution of financial market structure, where the correlation matrix of individual security returns is used as the primary item of study.\cite{Pan2007,Fenn2011,Mnnix2012,Heckens2020} Studying the nuanced connectedness between various groups of assets provides insights into the interdependencies of certain sectors, and how the market can be decomposed into smaller communities.\cite{Bonanno2003,Onnela2003,Onnela2004,Utsugi2004,Kim2005} In more recent years, network analysis has been used in various applications in financial markets, such as equities, \cite{Leibon2008} foreign exchange, \cite{Mikiewicz2021} cryptocurrency transactions, \cite{Kondor2014} mergers of companies, \cite{Fang2019_Matjaz} and global volatility. \cite{Cheng2022_volatility} Most of these analyses originally derive from the correlation matrix, but nonlinear alternatives also exist.\cite{Fiedor2014_2}

Like previous work, this paper focuses on identifying structurally important (and locally similar) communities of stocks, where each independent cluster on the network exhibits local affinity and dissimilarity with global comparisons across the graph. Our analysis also makes use of a modern and nuanced division of our universe of stocks into 60 sectors. The existing GICS (Global Industry Classification Standard) sectors are rather old, limited in number, fail to caption important differences in sector subdivisions, and have numerous companies that are misclassified by today's understanding. Our use of 60 sectors augments and aides the process of trying to find relevant market communities and themes.

We must also acknowledge the influence of statistical physics and time series analysis on this work, particularly Section \ref{sec:Market_structure_shifts}. In financial markets, time series analysis has been applied to a wide range of asset classes such as equities, \cite{james2022_stagflation,Wilcox2007,Alves2020,James2023_financialcrises} foreign exchange, \cite{Ausloos2000} cryptocurrencies, \cite{Gbarowski2019,james2021_crypto2,DrodKwapie2022_crypto,DrodWtorek2022_crypto,Drod2020,James2023_cryptoGeorg,Drod2023_crypto2,DrodWtorek2023_crypto} and debt-related instruments. \cite{Driessen2003} Such methods have been used more broadly in a variety of other disciplines including epidemiology, \cite{jamescovideu,Manchein2020,Li2021_Matjaz,Blasius2020,james2021_TVO,Perc2020,Machado2020,james2021_CovidIndia,james2023_covidinfectivity,Sunahara2023_Matjaz} environmental sciences, \cite{james2022_CO2,Khan2020,Derwent1995,james2021_hydrogen,Westmoreland2007,james2020_Lp,Grange2018,james2023_hydrogen2,Libiseller2005} crime, \cite{james2022_guns,Perc2013,james2023_terrorist} and other fields \cite{Clauset2015,james2021_olympics}.

Traditionally, the study of time-varying market shifts has rested on regime-switching models. These begin with the observation that financial assets exhibit  switching patterns, moving between periods of heightened volatility and ease. \citep{Hamilton1989,Lavielle2007,Lamoureux1990} One then applies the analysis of locally stationary time series \citep{Priestley1965,Priestley1973,OzakiTong,Tong1980} to partition a long time window into designated and discrete segments. These segments are classified into a small number of distinct regimes, usually two, \citep{Taylor1999,Taylor2005,Guidolin2011} but often different numbers \cite{Baba2011} or with varying assumptions that may yield different levels of flexibility. \citep{Arouri2016,Song2016,Balcombe2017,james_arjun,Carstensen2020,CerboniBaiardi2020,Campani2021} In addition, change point or structural break detection algorithms have been applied for a similar purpose, \cite{James2020_nsm,Ross2013physa,james2021_MJW} to divide a time interval into periods of fundamentally distinct market behavior. These do not have a specified number or type of regime in mind, nor do they classify and group different segments, but they do divide the entire window of analysis into discrete periods. While these discretized divisions have their place, they are a little contrived, so we instead develop a suite of different methods that reveal market shifts on a continual basis, and interpret their different uses and benefits.

The study of portfolio optimization has grown widely in the decades since Markowitz' mean-variance model. \cite{Markowitz1952,Sharpe1966} While there are many approaches including statistical mechanics, \cite{Zhao2016,Li2021_portfolio,james2021_portfolio} clustering, \cite{Iorio2018,Len2017} fuzzy sets, \cite{Tanaka2000,Ammar2003}  regularization, \cite{Fastrich2014,Li2015,Pun2019} Bayesian approaches \cite{james2021_spectral} and multiobjective optimization, \cite{Lam2021} almost all of these approaches aim to find a unique optimal portfolio. Frequently, NP hard constraints \cite{Shaw2008,Jin2016} such as portfolio cardinality constraints \cite{Anagnostopoulos2011} are imposed, which can make the selection of a single portfolio a difficult computational problem. In addition to being computationally difficult, finding a single optimal portfolio may be sensitive to changes in the data and may prove inflexible for future decision making. Thus, in this paper, we conduct a series of appropriately chosen sampling experiments to instead identify collections (specifically quantiles) of top performing portfolios and analyze their composition.

This paper is structured to address the broad topic of dynamic asset allocation and optimal portfolio construction during changing market conditions. Our paper proposes new mathematical techniques to address the three most critical aspects of tactical asset allocation: identifying market regimes, highlighting association between underlying portfolio components (assets, trading strategies, and so on) and optimally constructing portfolios within a sample space (with some degrees of freedom). In Section \ref{sec:Market_structure_shifts}, we compare a variety of new techniques to identify shifts in behavior in the market, consistently taking into account the structure of the market divided into our 60 sectors. We also contrast the findings and effectiveness of these techniques in various portfolio management contexts. In Section \ref{sec:Network_structure}, we turn to the network structure of the market and use our nuanced 60 sectors and network analysis to identify key clusters of structurally similar equity themes, and test for essential ingredients in long-term portfolio diversification. In Section \ref{sec:Portfolio_sampling}, we conduct an extensive sampling experiment where we compare the long-run performance and portfolio composition of long-only, short-only and long-short portfolios, taking into account the sector structure of the market.  In Section \ref{sec:conclusion}, we conclude. Throughout the paper, we take an interdisciplinary approach, as often seen in econophysics, borrowing methodologies from a wide range of mathematical foundations.

\section{Data}
\label{sec:data}

We analyze the daily closing prices of 268 equities between 2005-01-01 and 2023-12-31. Each equity has been allocated into one of 60 (new) sectors that more closely tie with their respective companies' roles in the business cycle. We feel that our nuanced sector allocation, which is much richer than the traditional Global Industry Classification Standard (GICS) methodology, is key in producing high quality insights. These sector allocations are done manually based on market consensus and are provided in Appendix \ref{app:sectorslist}. Price data has been sourced from Yahoo Finance.\cite{Yahoo_finance}

\section{Market structure shifts}
\label{sec:Market_structure_shifts}

The primary objects of study in this section are aggregated sector log returns (averaged over their composite assets) investigated over 30-day periods. We present four primary measures of deviation to identify temporal shifts in market structure. As we consistently group returns by sector, such changes may identify dislocations in sector pricing and hence arbitrage opportunities.

As mentioned in Section \ref{sec:data}, the full 268 stocks are divided into $n=60$ detailed sectors, $S_j, j=1,...,n$. Our data consists of asset prices from 2005-01-01 to 2023-12-31. We index the trading days as $t=0,1,...,T$ with $T=4780$. Averaged daily sector returns are defined as follows:

\begin{align}
    \label{eq:sectorreturn}
    R_j(t) = \frac{1}{|S_j|} \sum_{i \in S_j} \log \left( \frac{P_i(t)}{P_i(t-1)} \right), t=1,...,T.
\end{align}
Concretely, this represents the return of a portfolio that is concentrated entirely within one sector and equally weighted on the assets therein. Next, we consider $\tau=30$-day periods of returns, which we donate $R_j[t-\tau+1:t], t=\tau,...,T$. This is to be understood as a sequence of adjacent sector returns over 30 days.

This analysis of 30-day intervals requires comment, specifically around the choice of a 30-day window and alternatives. In general, smoothing out a time series is a balance with respect to signal and noise. Given the high level of noise in financial time series, de-noising these time series often helps us make better predictions as we can more accurately capture the true "signal" in the underlying time series. A 30-day window acts as a low-pass filter; a larger window induces more bias into the estimator but reduces the variance more substantially, while a smaller window yields less bias. In our previous work, we have experimented with several time lags, windows, offsets, and so on. Having applied similar methods to underlying financial time series, a 30-day window tends to generalize best. As we acknowledge in Section \ref{sec:conclusion}, alternative time windows may be an avenue for future work.

\begin{figure}
    \centering    
    \includegraphics[width=0.49\textwidth]{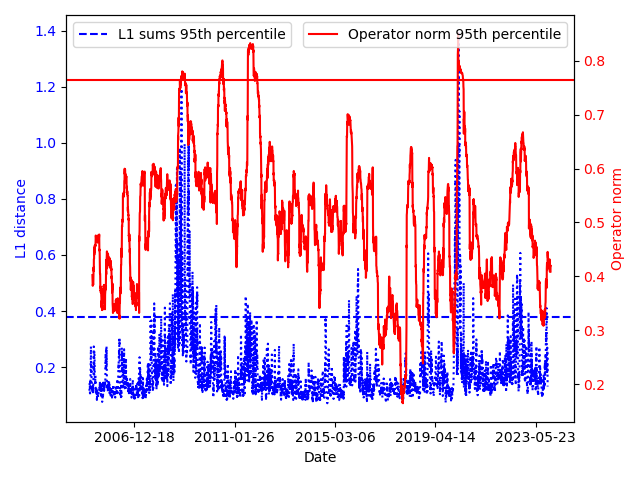}
    \caption{Line plots of $S_t$ (adjacent $L^1$ sums between sector returns) and $C_t$ (normalized operator norms of rolling correlation matrices), defined in (\ref{eq:distance1}) and (\ref{eq:lambda1}), respectively. Both estimators provide consistent indication of financial crisis detection, with $S_t$ exhibiting greater variability. We include thresholds noting the top 5\% of respective values of each series. There is no relationship between the two $y$-axis scales.}
    \label{fig:L1_lambda_plot}
\end{figure}

First, we sum over the 30-day sequences $R_j[t-\tau+1:t]$ to yield monthly returns. That is, let $\bar{R}_j[t-\tau+1:t] = \sum R_j[t-\tau+1:t] = \sum_{t-\tau+1 \leq s \leq t} R_j(s)$. With this, our first measure of shifts in market structure is defined as
\begin{align}
\label{eq:distance1}
S_t = \| \bar{\mathbf{R}}[t-\tau+1:t] - \bar{\mathbf{R}}[t+1:t+\tau] \|_1 \\ = \sum_{j=1}^n \left| \bar{R}_j[t-\tau+1:t] - \bar{R}_j[t+1:t+\tau] \right| \\ = \sum_{j=1}^n \left| \sum_{t-\tau+1 \leq s \leq t} R_j(s) - \sum_{t+1 \leq s \leq t+\tau} R_j(s) \right|.
\end{align}
Above, $\bar{R}_j$ denotes averages taken over time, while $\bar{\mathbf{R}}$ denotes the length $n=60$ vector of all $\bar{R}_j$ values across all sectors, $j=1,...,n$. The $\| \cdot \|_1$ notation signifies an $L^1$ norm used as a metric between vectors. $S_t$ is defined for $t=\tau,...,T-\tau$ and computes the shift in returns' monthly \emph{sums} on a sector-by-sector basis between the 30-day periods immediately prior to and following $t$. We plot this in Figure \ref{fig:L1_lambda_plot}.

Our second measure is slightly more sophisticated. We retain $R_j[t-\tau+1:t]$ as sequences without summing them, and instead compute \emph{Wasserstein distances} between these sequences, understanding them as distributions. That is, let $d_W$ be the $L^1$ Wasserstein metric. Consider each length-30 sequence $R_j[t-\tau+1:t]$ of daily log returns as a distribution of real values. The Wasserstein metric $d_W(R_j[t-\tau+1:t],R_j[t+1:t+\tau])$ is defined informally as the work required to transform the distribution $R_j[t-\tau+1:t]$ into $R_j[t+1:t+\tau]$, and is outlined in Appendix \ref{app:wasserstein}.

The Wasserstein metric $d_W$ offers a way to simultaneously compute aggregated measures of disparity between two sequences (considered as distributions) while allowing individual terms to remain relevant. For example, a strong negative shock $R_j(t+1)$ followed by positive shock $R_j(t+2)$ in the return sequence $R_j[t+1:t+\tau]$ could ``cancel out'' in the computation of $S_t$, whereas both terms will remain relevant in the computation of $W_t$ below.

\begin{figure}
    \centering    
    \includegraphics[width=0.49\textwidth]{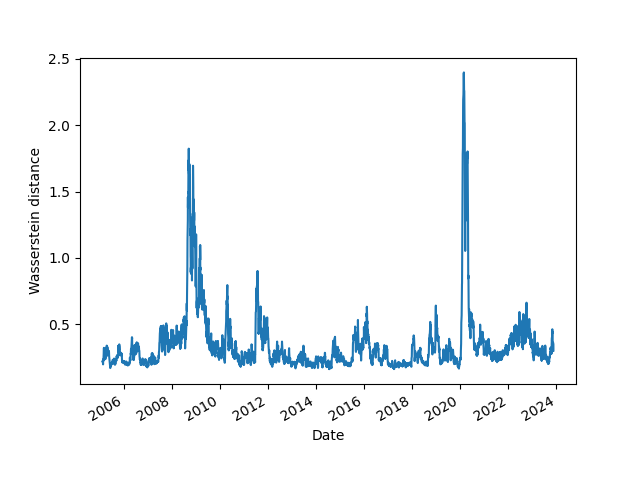}
    \caption{Line plot of $W_t$, defined in (\ref{eq:distance2}). These Wasserstein distances provide a more abrupt estimator, which may exhibit a higher signal-to-noise ratio for short-term anomaly identification. The metric identifies the cumulative deviation in sector-to-sector log returns distributional changes. Large measurements are likely associated with latent structural shifts in the market. }
    \label{fig:Wasserstein}
\end{figure}

With this, our second measure of shifts in market structure is defined as
\begin{align}
\label{eq:distance2}
    W_t = \sum_{j=1}^n d_W \left(R_j[t-\tau+1:t], R_j[t+1:t+\tau] \right),
\end{align}
again defined for $t=\tau,...,T-\tau$. This computes the shift in returns' monthly \emph{distributions} on a sector-by-sector basis between the 30-day periods immediately prior to and following $t$. It is plotted in Figure \ref{fig:Wasserstein}. 

Our third measure of shifts in market structure reflects the collective strength of correlations between sector returns, computed over a rolling 60-day basis. Specifically, for $j,k=1,...,n$ and $t=\tau,...,T-\tau$, we define
\begin{align}
\label{eq:corrmatrix}
\Psi_{jk}(t)= \text{Corr}(R_j[t-\tau+1:t+\tau], R_k[t-\tau+1:t+\tau]), 
\end{align}
where Corr is the Pearson correlation between length-60 vectors. For each $t$, we obtain a $n \times n$ matrix $\Psi(t)$ that is symmetric with real eigenvalues $\lambda_1(t) \geq ... \geq \lambda_n(t) \geq 0$. All diagonal elements of $\Psi(t)$ equal 1. Thus, the trace of the matrix, and hence the sum of the eigenvalues, equals $n$. We normalize the leading eigenvalue by dividing by $n$ and define $C_t = \frac{\lambda_1(t)}{n}.$ This computes the collective strength of correlations between sectors across both 30-day periods immediately prior to and following $t$, concatenated into a 60-day period. In addition, $\lambda_1(t)$ can be interpreted as the \emph{operator norm} of $\Psi(t)$:
\begin{align}
\label{eq:lambda1}
    \lambda_1(t) = \max_{v \in \mathbb{R}^n} \frac{\|\Psi(t)v \|}{\|v\|}.
\end{align}
Informally, one can think of this measure as the maximum factor by which a vector is lengthened under a linear map. We plot $C_t$ in Figure \ref{fig:L1_lambda_plot} to compare with $S_t$.

\begin{figure*}
    \centering
    \begin{subfigure}[b]{0.49\textwidth}
        \includegraphics[width=\textwidth]{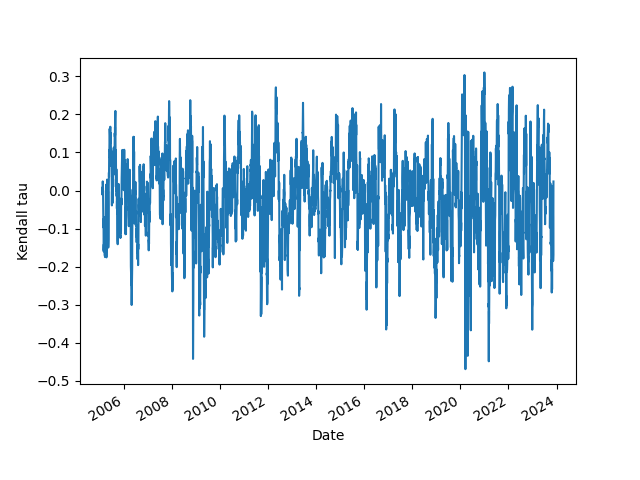}
        \caption{}
        \label{fig:Kendall}
    \end{subfigure}
    \begin{subfigure}[b]{0.49\textwidth}
        \includegraphics[width=\textwidth]{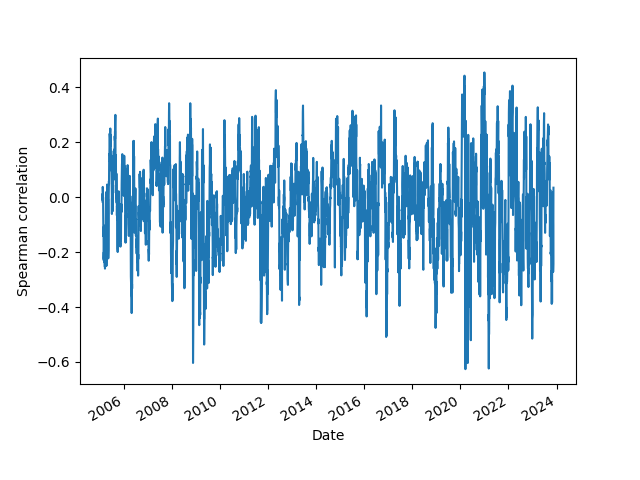}
        \caption{}
        \label{fig:Spearman}
    \end{subfigure}
    \begin{subfigure}[b]{0.49\textwidth}
        \includegraphics[width=\textwidth]{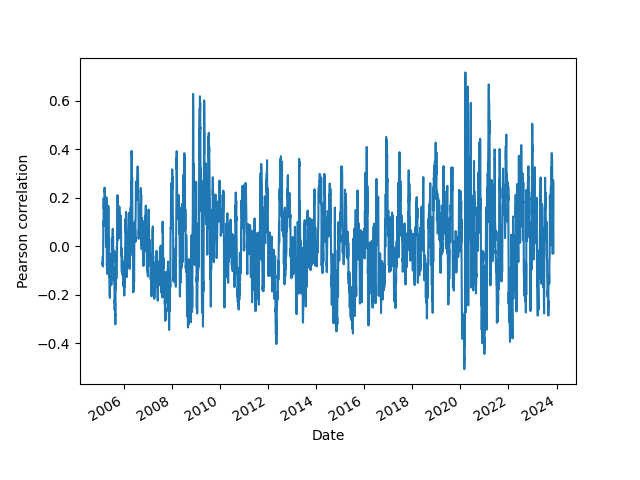}
        \caption{}
        \label{fig:pearson}
    \end{subfigure}
    \begin{subfigure}[b]{0.49\textwidth}
        \includegraphics[width=\textwidth]{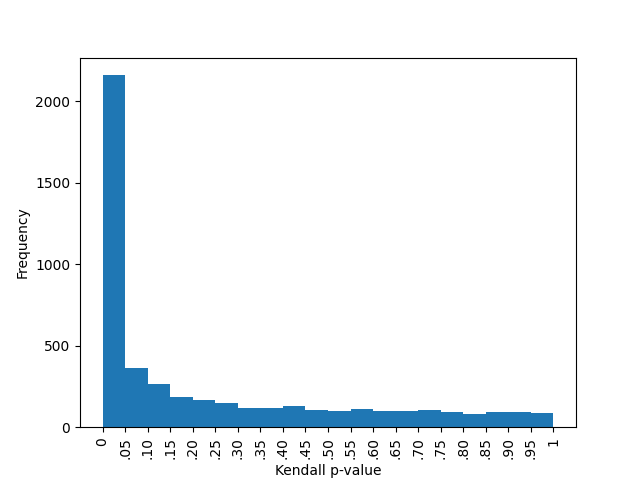}
        \caption{}
        \label{fig:kendallpval}
    \end{subfigure}
    \caption{In (a), we display the rolling Kendall tau $K_t$ over time, defined in Section \ref{sec:Market_structure_shifts}. For robustness, we include the closely related Spearman correlation (b) and more traditional Pearson correlation (c). The Kendall and Spearman are determined solely in terms of changing ranks of sector log returns. In (d), we include the $p$-values associated with the Kendall tau computations. The more robust Kendall and Spearman coefficients show a consistently changing order in sector return performance on a rolling basis, with little change to this observation during the GFC or COVID-19 crises.}
   \label{fig:Kendall_spearman_plots}
\end{figure*}

Finally, we include one additional measure of deviation that targets not just magnitudes of deviations in returns between 30-day periods, but changes in the ranking order of sector returns. Once again, we utilize the 30-day summed returns (equivalently monthly log returns) $\bar{R}_j[t-\tau+1:t]$. Let $K_t$ be the \emph{Kendall tau} coefficient between the two vectors $(\bar{R}_j[t-\tau+1:t])_j$ and $(\bar{R}_j[t+1:t+\tau])_j$ in $\mathbb{R}^n$, where $j$ ranges over $1,...,n=60$ and produces two vectors of dimension 60. The Kendall tau coefficient is computed as follows: a pair $1\leq i <j \leq n$ is said to be concordant between the two vectors if the order of $(\bar{R}_i[t-\tau+1:t], \bar{R}_j[t-\tau+1:t] )$ and $(\bar{R}_i[t+1:t+\tau], \bar{R}_j[t+1:t+\tau])$ are the same, and discordant otherwise. Then the coefficient is defined as the difference between the number of concordant and discordant pairs as a proportion of all ${n \choose 2}$ pairs. The Kendall tau has an associated $p$-value corresponding to the null hypothesis of a zero tau coefficient between vectors. We plot $K_t$ in Figure \ref{fig:Kendall_spearman_plots}, again for $t=\tau,...,T-\tau$. For sake of robustness, we also include the related Pearson and Spearman correlation coefficients between the same vectors. The Spearman correlation is defined as the Pearson correlation computed between the ranks of the two vectors, and is thus computed entirely based on the ranks of the sector return values, not their relative magnitudes.

Figure \ref{fig:L1_lambda_plot} reveals some key similarities between our measures $S_t$ and $C_t$. First, both measures identify the global financial crisis (GFC) and COVID-19 market crash as the most significant equity market crises over the full period investigated. We consider values in the top 5th percentile of each respective series as indicative of the most atypical market dynamics and mark these thresholds in the figure. Both measures detect three breaches of this 5\% threshold during the GFC and one during COVID-19. Further, both measures broadly correspond in amplitude over the entirety of the period studied: generally, they are relatively high and low during the same times. 

And yet, some differences between the quantities exist that distinguish them as estimators of structural market shifts. First,  $S_t$ may serve as a more effective tool in an online capacity for market turmoil identification. Prior to the GFC and COVID-19 market crash, smaller scale perturbations are detected, which could be used as an indicator for subsequent sell-offs in equity markets. Next, $S_t$ produces greater contrast in magnitude when compared to the $C_t$ measurement. This is highlighted during the GFC: both $C_t$ and $S_t$ produce three spikes at approximately the same time, but the spikes of $S_t$ exhibit greater variability in amplitude. On the other hand, one potential drawback of the $S_t$ measurement is the potential for false positive structural shift identification. During the period investigated, 11 breaches of the 5\% threshold are identified using $S_t$. By contrast, $C_t$ only identifies four structural breaks - three during the GFC and one during the COVID-19 market crisis.

Our more refined measure for time-varying market shifts is shown in Figure \ref{fig:Wasserstein}, and determines the rolling, cumulative sector-to-sector Wasserstein distance between adjacent length-30 sequences of log return distributions. Figure \ref{fig:Wasserstein} is quite different from the three other measures we have employed, for several reasons. First, our Wasserstein metric approach produces the most abrupt changes around market crises. One could argue that this methodology has a higher signal-to-noise ratio with respect to crisis identification, and given the strong association between the amplitude of the estimator and the actual existence of financial market crises - it may indeed be the most effective tool for regime or crisis identification. This is to be expected, as using the Wasserstein metric rather than the sum in $S_t$ allows one to more carefully take into account extreme daily values, even if a sharp positive market shift is directly followed by a sharp negative shift. Indeed, such behavior is quite common during crises, and only using the Wasserstein metric will properly detect these unique patterns. Like our $S_t$ methodology, the Wasserstein approach produces significant variability in amplitude. By contrast, $C_t$, which captures the evolutionary change in the amount of collective market movement, does not exhibit the same variance in amplitude. This framework may have favourable properties for online regime identification, crisis detection and be useful as an indicator of market dislocations or required shifts in optimal portfolio composition.

Finally, we display the Kendall tau $K_t$ in Figure \ref{fig:Kendall}, along with the closely related Spearman correlation coefficient in \ref{fig:Spearman} and the more standard Pearson coefficient in \ref{fig:pearson}. Unlike our other time-varying measures, Figure \ref{fig:Kendall}  displays the evolution of change in sector log returns ranks, and perhaps best indicates the need for changes in portfolio composition over time. The three figures display some similarity in their evolutionary structure and volatility. However, it is notable that pronounced spikes around the GFC and COVID-19 crash are only clearly visible in the Pearson figure \ref{fig:pearson} and not the others. Log returns are not normally distributed, particularly during market crises, and so the Pearson coefficient may be distorted by extreme values during these times. Thus, looking only at the ranks (Kendall tau) or correlations between the ranks (Spearman) reveals no particular shifts in rank orderings during the crises. Instead, we see considerable consistency over time in the behaviors of $K_t$ and the closely related Spearman coefficient, included for robustness. The magnitudes of the Kendall tau and Spearman coefficient are generally quite low, never exceeding 0.3 and 0.4 respectively, and more frequently close to 0. To complement this, we include a histogram of the associated Kendall tau p-values in Figure \ref{fig:kendallpval}. Less than half (of 4721) of the $p$-values are less than 0.05, with many distributed between 0.05 and 1, indicating that there are many times $t$ where the Kendall tau $K_t$ is not significantly different from zero. This indicates a considerable shift in ordering from one rolling period of returns to another. Both of the above findings indicate that optimal portfolios must switch between market sectors on a near-continual basis.

\section{Network market structure}
\label{sec:Network_structure}

Let $\Psi_{ij}$ be the $n \times n$ correlation matrix between sectors calculated over the entire period of analysis. This is defined analogously to the $n \times n$ matrices $\Psi(t)$ in (\ref{eq:corrmatrix}) but over the entire interval $[1,T]$ rather than a 60-day window. We apply the following common transformation:
\begin{align}
\label{eq:transformcorr}
D_{ij} = \left(2(1-\Psi_{ij}) \right)^\frac{1}{2}.
\end{align}
This is motivated by the fact that correlation can be interpreted as a normalized inner product, or an inner product between unit vectors $\langle u, v \rangle $. For unit vectors $u,v$, we have the following identity:
\begin{align}
\label{eq:innerproduct}
\| u - v \|^2 = \langle u, u \rangle + \langle v, v \rangle - 2 \langle u, v \rangle = 2(1-\langle u, v \rangle ),
\end{align}
where $\| \|$ is the norm associated to the inner product $\langle ., . \rangle $. Thus, the transformation (\ref{eq:transformcorr}) yields the metric between unit vectors that is associated to correlation.

We then form two networks, one traditional and one somewhat unusual. First, more traditionally, we form a weighted graph using the $n$ sectors as vertices and the associated distances $D_{ij}$ as edge weights. In this graph, two vertices (sectors) have a small edge weight between them if they are highly correlated (high correlation being equivalent to small $D_{ij}$). Then, we pass this to Kruskal's algorithm to form an associated minimum spanning tree (MST). 

On the other hand, a little unusually, we form a weighted graph directly using the correlation coefficients $\Psi_{ij}$ themselves as edge weights. In this graph, two vertices are close if they have a low correlation. Unlike $D_{ij}$, $\Psi_{ij}$ does not satisfy the triangle inequality property of a metric, so there is no transitive property of closeness. However, we once again form a minimum spanning tree on this weighted graph. This tree will reveal the minimal sequence of edges to connect the graph subject to wanting connected sectors to be uncorrelated, which is important in diversification.

Kruskal's algorithm to produce a minimum spanning tree proceeds as follows. In a weighted graph, one first orders the edges by increasing weight. Then, edges are successively added to the tree if and only if the addition of such an edge would not create a cycle. By design, the algorithm never produces any cycles, but successively adds edges until the graph is connected, producing a connected spanning graph with no cycles, and hence a tree. We remark that Kruskal's algorithm works fine when some edges of the graph have negative weight, which may be the case for the correlation weightings $\Psi_{ij}$. The resulting MST still aims to minimize path lengths with respect to correlation; negative correlation values are fine and indeed are even desired to show assets with low correlation, which is beneficial for diversification. Concretely, Kruskal's algorithm finds the exact same MST as if we adjusted all edges by a constant, for example examining edges $\Psi_{ij}+1$, all of which would be non-negative.

\begin{figure*}
    \centering
    \includegraphics[width=\textwidth]{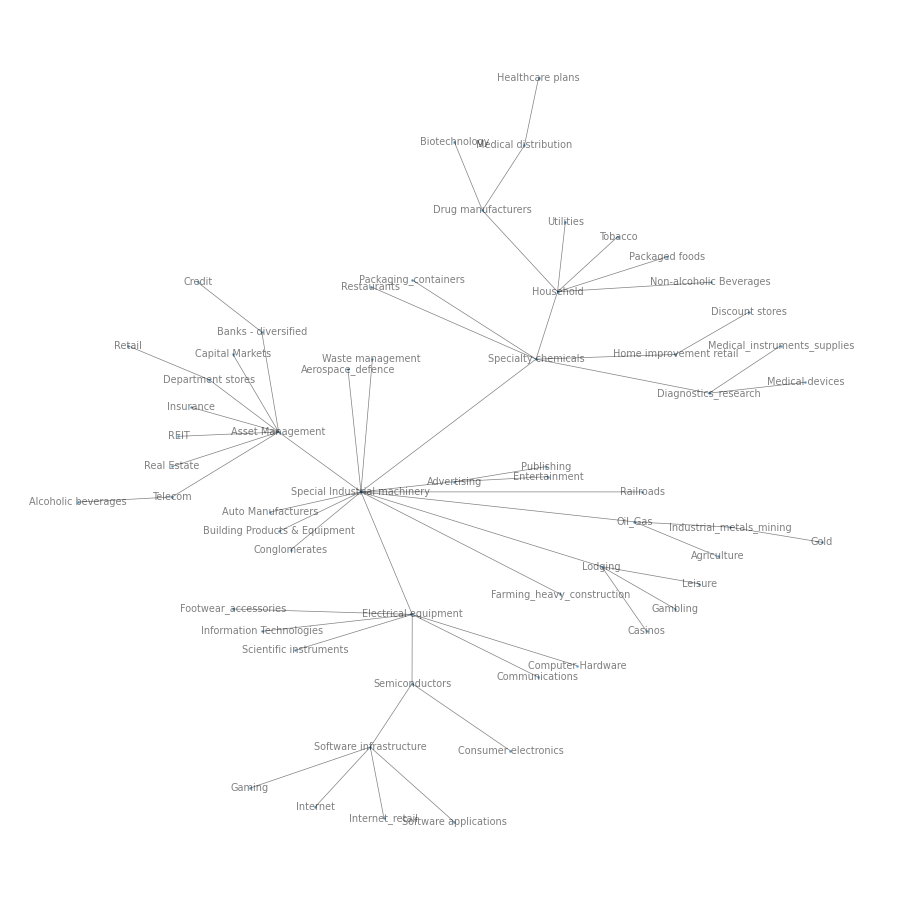}
    \caption{Minimum spanning tree associated to the distance matrix $D$ indicating high similarity between sectors. We see numerous themes in market structure, including manufacturing (more central), electronics/computer manufacturing and software (to the bottom), household (top right), financial services (left).}
    \label{fig:MST Metric full period}
\end{figure*}

\begin{figure*}
    \centering
        \includegraphics[width=\textwidth]{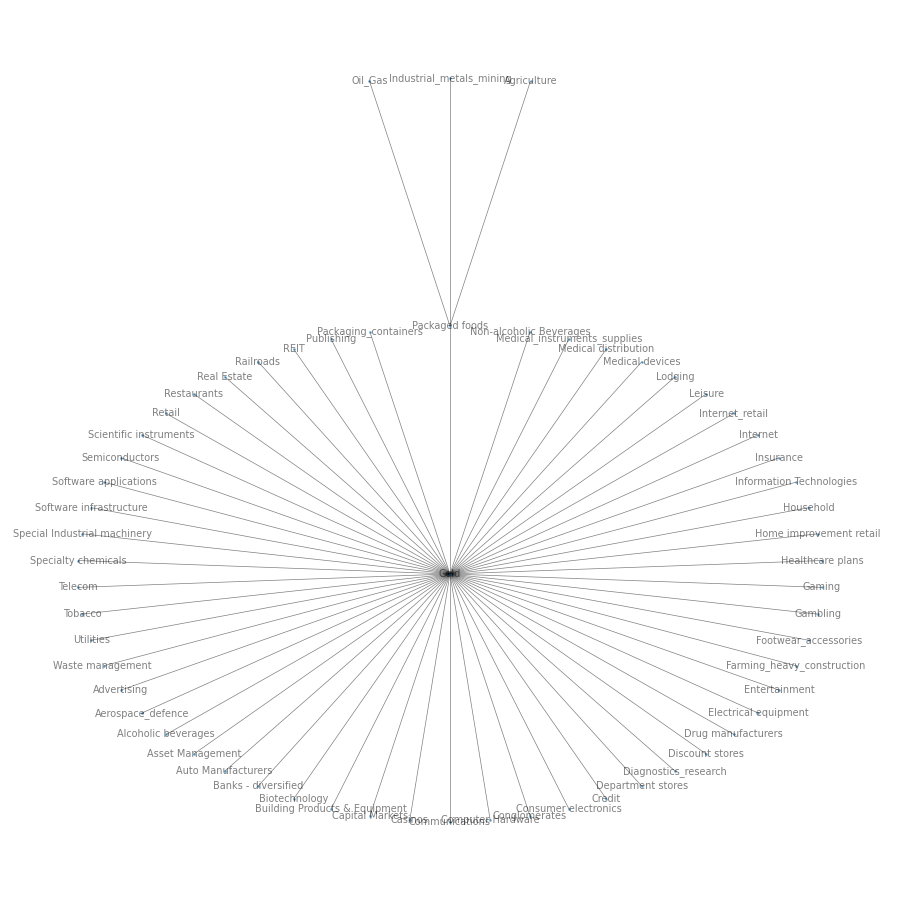}
        \caption{Minimum spanning tree directly associated to the correlation matrix $\Psi$, where edges link sectors of low correlation. This showcases the central role of the gold sector as the asset class with the least correlation with almost every other sector.}
        \label{fig:MST Correlation full period}
\end{figure*}

In Figures \ref{fig:MST Metric full period} and \ref{fig:MST Correlation full period}, we display the minimum spanning tree using $D$ and $\Psi$ as the edge weights between sectors considered as vertices, respectively. In the first network, Figure \ref{fig:MST Metric full period}, we identify sectors that are more closely correlated. Several noteworthy clusters are identified, and such groupings could be used to identify broad themes among the market. First, we see a strong association between software applications, internet and software infrastructure companies. These sectors are connected to semiconductors and consumer electronics, with the latter two exhibiting higher centrality on the graph. These nodes are connected to sectors such as computer hardware, information technologies, electrical equipment and scientific instruments, which are again more central to the underlying graph. This entire collection of sectors could form a primary "digitization" theme, or be broken down into more detailed sub-themes such as "software businesses" and "computing infrastructure". We also see strong association between healthcare plans, medical distribution, biotechnology and drug manufacturers, forming a "healthcare" theme. One can also see a collection of sectors that includes discount stores, home improvement retail, packaging containers and restaurants, forming a theme related to "consumer goods and services." The groupings that are identified on the graph could be used in several contexts. One potential application would be forming new communities of equities based on the exposed themes, and re-computing sector attributes such as expected returns, volatilities and beta against the larger market. One could use these themes as an alternative to portfolio factors, or could diversify a portfolio with respect to these algorithmically-determined themes (rather than GICS sectors).

In Figure \ref{fig:MST Correlation full period}, we display the minimum spanning tree associated to the correlation matrix $\Psi$, directly. Edges are drawn between equities with lower correlation, rather than greater similarity. The structure of this network is striking, almost entirely centered around the gold sector (which contains four assets) as a star-shaped graph. Almost every sector shares an edge only with gold, because the gold sector repeatedly exhibits the lowest pairwise correlation relative to anything else. The centrality and importance of the gold sector demonstrates that gold (and gold-related businesses) is a vital asset in diversifying financial portfolios in all weather settings.

\section{Portfolio sampling}
\label{sec:Portfolio_sampling}

\subsection{Sampling methodology}

In this section, we conduct a series of simulation studies across different sample spaces. We sample over long-only, short-only and long-short portfolios, uniformly across all stocks or stratified by sectors. This allows us to investigate the composition of strong performing portfolios across sectors and identify sectors that are disproportionately represented in top performing portfolios. We complement previous work that only varied portfolio size as the determinant of the sample space and did not consider any sector structure. \cite{James2024_portfolioEPL}

Let $\mathcal{S}_L$ be the sample space consisting of all long positions in the $N=268$ equities. Symmetrically, let $\mathcal{S}_S$ be the sample space consisting of the corresponding short positions. These produce the exact negative returns of the long positions, thus
\begin{align}
\label{eq:shortvlong}
\mathcal{S}_S= - \mathcal{S}_L.
\end{align}
In our experiments, we primarily sample portfolios of size $m=30$ from both $\mathcal{S}_L$ and $\mathcal{S}_S$. In light of common investor portfolio statements that impose tight constraints on the weights of individual assets \cite{Russellpolicy} and research indicating it is frequently difficult to beat equally weighted portfolios, \cite{DeMiguel2007,Farago2022} we restrict our sampling to equally weighted portfolios. Thus, a sampled long-only or short-only portfolio of size $m$ from our universe of equities is equivalent to a sample drawn of size $m$ from $\mathcal{S}_L$ or $\mathcal{S}_L$ respectively.

Next, let $\mathcal{S}_{SL}$ be the sample space consisting of all long and/or short positions in the $N=268$ equities. There is a subtle distinction that yields two different ways to define this sample space. First, perhaps more naively, we can interpret $\mathcal{S}_{SL}$ as a disjoint union
\begin{align}
\label{eq:shortlongunion}
\mathcal{S}_{SL} = \mathcal{S}_{L} \cup -\mathcal{S}_{L} \\ = \{x: x \in  \mathcal{S}_{L}\} \cup \{- x: x \in  \mathcal{S}_{L}\}.
\end{align}
Then, a sampled long-short portfolio of size $m$ is equivalent to a sample drawn of size $m$ from $\mathcal{S}_{SL}$ above. However, such a sampled portfolio may contain both a long and a short position of the exact same asset, which would be equivalent to holding neither. This is not necessarily entirely redundant, as it is common for investment teams to hedge one another, simultaneously taking long and short positions of similar or identical assets. For robustness, we also present an alternative sample space that explicitly prohibits a long and short position in the same equity. Consider the Cartesian product
\begin{align}
\label{eq:shortlongunion2}
\mathcal{S}_{SL}^{(m)} = \mathcal{S}_{L}^m \times \{1, -1\}^m.
\end{align}
Then, a long-short portfolio of size $m$ can also be understood as a pair of samples, one $m$-length sample from $\mathcal{S}_{L}$ and one $m$-length sample from $\{1, -1\}$. Then, if $A_1,...,A_m$ are $m$ distinct assets, we write any long-short portfolio with no duplicated long and short positions as
\begin{align}
    \frac{1}{m}\left( \epsilon_1 A_1 + ... + \epsilon_m A_m \right)
\end{align}
where $\epsilon_i \in \{1, -1\}$; then no asset can be held both long and short simultaneously. In all experiments, we observe no discernable difference in results between the two long-short sample spaces, so we proceed with the second definition.

With these three sample spaces in mind, we now present one further distinction. While we always define a portfolio using equal weights, we conduct two distinct experiments regarding the random drawing of assets from $\mathcal{S}_{L}$, $\mathcal{S}_{S}$ and $\mathcal{S}_{SL}^{(m)}$. In one set of experiments, which we term \emph{uniform sampling}, we sample any asset with equal probability. That is, each of the $N=268$ equities may be chosen with equal probability from the sample space $\mathcal{S}_{L}$. In another set of experiments, which we term \emph{stratified sampling}, we sample each equity with a probability that is inversely proportional to the size of the sector. Specifically, let the $n=60$ sectors have sizes $k_1,...,k_n$. Then an equity in sector $i$ is sampled with probability
\begin{align}
    p_i=\frac{1}{nm_i}=\frac{1}{60k_i}. 
\end{align}
Then for any sector, the total probability of an asset in that sector being chosen is equal to $\frac{1}{60}$. This ensures each sector is represented the same  number of times (on average) throughout our sample draws, effectively adjusting uniform sampling for sector size.

In all these sampling experiments, our primary quantity of interest is the Sharpe ratio of a portfolio, defined as 
\begin{align}
\label{eq:Sharpeobjectionfn}
 \frac{\sum^{k}_{i=1} w_{i} R_{i} - R_f}{ \sqrt{\boldsymbol{w}^{T} \Sigma \boldsymbol{w}}  }.
\end{align}
In (\ref{eq:Sharpeobjectionfn}) above, $R_i$ is the return of a stock over the entire period, $\Sigma$ is the covariance matrix between stocks over the period, $w_i$ is each stocks' portfolio weight, and $R_f$ is the risk-free rate. In this paper, we only use equally weighted portfolios, with all $w_i$ set to $\frac{1}{m}$ for a size $m$ portfolio, and a risk-free rate of 0. The Sharpe ratio is a measure of the risk-adjusted returns of a portfolio that seeks to reward portfolios with high returns, while simultaneously penalising excessive variance.

In Section \ref{sec:samplingresults}, we report numerous results from sampling both uniform and stratified portfolios, either long-only, short-only or long-short, and for various values of portfolio size $m$, most commonly $m=30$. In every instance, we draw 10 000 samples from each sample space in each experiment. We report both quantiles of the annualized Sharpe ratio across the entire sample space, as well as the composition of portfolios by sector in top 1\% performing portfolios, quantified and ordered by Sharpe ratio.

\subsection{Sampling results}
\label{sec:samplingresults}

Table \ref{tab:Sharpe_composition_table} displays Sharpe ratio quantiles for portfolios of size $m=10, 20, 30$ with uniform sampling and then $m=30$ with stratified sampling. A short-only trading strategy over such a prolonged period in the market, where equities exhibit survivorship bias and significantly positive cumulative returns, would almost certainly be a losing investment strategy. This is visible in the consistently negative values of the short-only Sharpe ratios. Accordingly, we do not focus on the short-only strategy and contrast the return profiles of the long-only and long-short strategies. The results demonstrate that the long-only portfolio is superior to the long-short portfolio in almost all points of the distribution. This highlights, over the long-term, the degree of manager skill that is required when incorporating more shorting into a portfolio. Our results indicate that among our collection of equities, short positions are best used when highly targeted and over discrete periods with finite end dates. Indeed, shorting allows limitless downside on portfolio returns. Furthermore, our simulations do not consider transaction costs, stock loan or margin requirements, which all add to the difficulty of beating a standard long-only portfolio. The inherent asymmetry associated to short positions further adds to the difficulty in long-short managers beating traditional long-only investors. The four parts of the table show that, predictably, the spread of quantiles narrows from $m=10$ to  20 and 30. In addition, a comparison of the third and fourth quarter of the table show essentially no difference in Sharpe distribution between stratified and uniform sampling.



\begin{table}[htbp]
\centering
\begin{tabular}{lrrr}
\toprule
Quantile &  Long-only &  Short-only &  Long-short \\
\midrule
$m=10$ & Uniform sampling & & \\
\midrule
0.01 &       0.23 &       -0.69 &       -0.66 \\
0.05 &       0.29 &       -0.62 &       -0.50 \\
0.10 &       0.32 &       -0.58 &       -0.41 \\
0.25 &       0.38 &       -0.52 &       -0.24 \\
0.50 &       0.45 &       -0.45 &        0.01 \\
0.75 &       0.52 &       -0.38 &        0.24 \\
0.90 &       0.58 &       -0.32 &        0.41 \\
0.95 &       0.61 &       -0.29 &        0.50 \\
0.99 &       0.68 &       -0.22 &        0.65 \\
\midrule
$m=20$ & Uniform sampling & & \\
\midrule
0.01 &       0.31 &       -0.64 &       -0.65 \\
0.05 &       0.35 &       -0.58 &       -0.50 \\
0.10 &       0.37 &       -0.56 &       -0.42 \\
0.25 &       0.42 &       -0.51 &       -0.25 \\
0.50 &       0.46 &       -0.47 &       -0.00 \\
0.75 &       0.52 &       -0.42 &        0.24 \\
0.90 &       0.56 &       -0.38 &        0.41 \\
0.95 &       0.59 &       -0.35 &        0.50 \\
0.99 &       0.64 &       -0.31 &        0.65 \\
\midrule
$m=30$ & Uniform sampling & & \\
\midrule
0.01 &       0.34 &       -0.61 &       -0.64 \\
0.05 &       0.38 &       -0.57 &       -0.50 \\
0.10 &       0.40 &       -0.55 &       -0.42 \\
0.25 &       0.43 &       -0.51 &       -0.25 \\
0.50 &       0.47 &       -0.47 &        0.00 \\
0.75 &       0.51 &       -0.43 &        0.24 \\
0.90 &       0.55 &       -0.40 &        0.42 \\
0.95 &       0.57 &       -0.38 &        0.50 \\
0.99 &       0.61 &       -0.34 &        0.65 \\
\midrule
$m=30$ & Stratified sampling & & \\
\midrule
0.01 &       0.34 &       -0.62 &       -0.66 \\
0.05 &       0.38 &       -0.58 &       -0.50 \\
0.10 &       0.40 &       -0.55 &       -0.42 \\
0.25 &       0.44 &       -0.52 &       -0.24 \\
0.50 &       0.47 &       -0.48 &        0.00 \\
0.75 &       0.52 &       -0.44 &        0.24 \\
0.90 &       0.55 &       -0.40 &        0.41 \\
0.95 &       0.58 &       -0.38 &        0.50 \\
0.99 &       0.62 &       -0.34 &        0.65 \\
\bottomrule
\end{tabular}
\caption{Comparison of Sharpe ratio quantiles for long-only, short-only and long-short portfolios of cardinality 10, 20 and 30 (uniform sampling) and then size 30 (stratified sampling). As $m$ increases, the spread of values narrows. Uniform and stratified sampling exhibit negligible difference in values.}
\label{tab:Sharpe_composition_table}
\end{table}

\clearpage

\begin{figure}
    \centering
    \begin{subfigure}[b]{0.49\textwidth}
        \includegraphics[width=\textwidth]{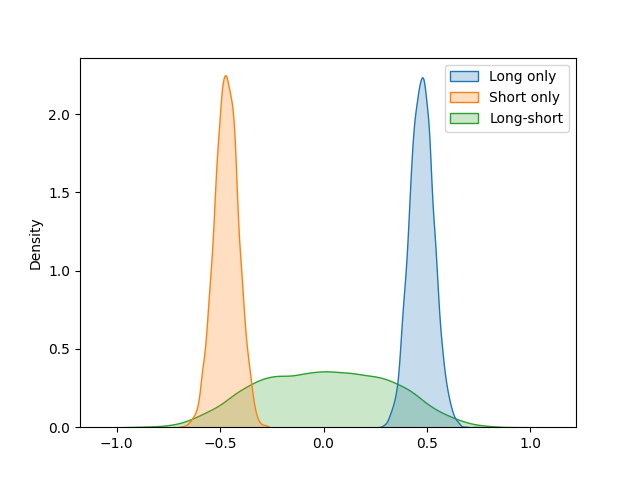}
        \caption{}
        \label{fig:Portfolio_sampling_30_density}
    \end{subfigure}
    \begin{subfigure}[b]{0.49\textwidth}
        \includegraphics[width=\textwidth]{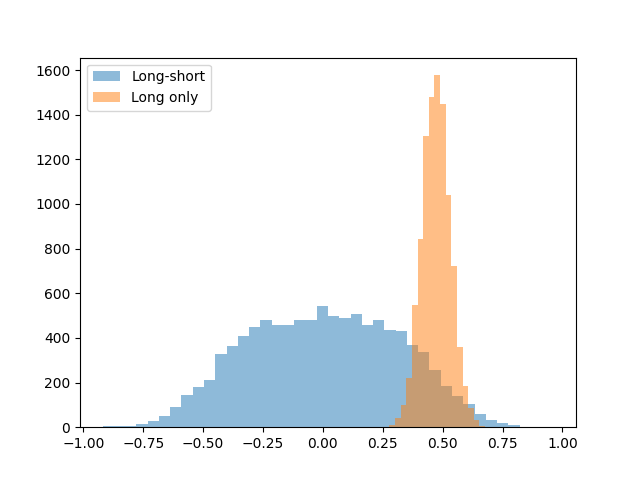}
        \caption{}
        \label{fig:Portfolio_sampling_30_hist}
    \end{subfigure}
    \begin{subfigure}[b]{0.49\textwidth}
        \includegraphics[width=\textwidth]{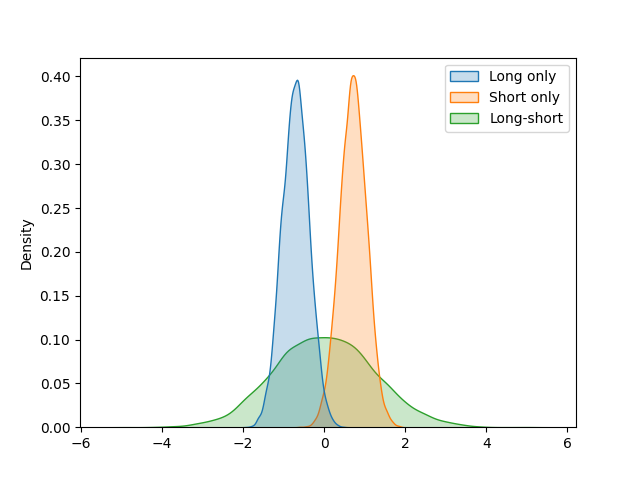}
        \caption{}
        \label{fig:Portfolio_sampling_30_density_GFC}
    \end{subfigure}
    \caption{Portfolio sampling for 30-stock portfolios. In (a) we show kernel density plots, (b) histograms and in (c) we show the contrast in results where we restrict data just to the GFC. There, we see almost the opposite result than over the entire period.}
   \label{fig:Portfolio_sampling}
\end{figure}

Figure \ref{fig:Portfolio_sampling} shows the kernel density estimation (\ref{fig:Portfolio_sampling_30_density})  and histogram (\ref{fig:Portfolio_sampling_30_hist}) of the Sharpe ratio distribution for long-short and long-only investment portfolios with $m=30$. Confirming Table \ref{tab:Sharpe_composition_table}, it is clear that long-only Sharpe ratios are superior in expectation to that of long-short portfolios due to the length of the period. We also visually see that the distribution of outcomes is much narrower for long-only Sharpe ratios - highlighting the increase in the size of the sample space when incorporating more degrees of freedom into the portfolio. However, the increase in the sample space is predominantly negative - further emphasizing the importance of manager skill when managing portfolios with more degrees of freedom. By way of contrast, Figure \ref{fig:Portfolio_sampling_30_density_GFC} repeats this experiment restricting to the period 2007-09-01 to 2009-03-31, comprising the global financial crisis (GFC). Here we see a clear shift, with short-only outperforming long-short and short-only. This reflects the fact that for investment managers to successfully engage in short-selling, they are better off doing so in focused bear market windows.

In Table \ref{tab:probcomparison}, we quantify this comparison between long-only and long-short distributions. Specifically, we compute $P(X>Y)$, the empirical probability that a randomly chosen long-short portfolio has a greater Sharpe ratio than a randomly chosen long-only portfolio. We compute this for $m=10, 20, 30$, using uniform and stratified sampling, and over the entire period or just the GFC. Confirming Figure \ref{fig:Portfolio_sampling}, we calculate a relatively small probability (around 7\%) that a candidate long-only portfolio across the entire period can beat long-short. However, this increases to around 70\% when restricting to the GFC. Curiously, we observe slightly smaller probabilities for stratified sampling than uniform sampling, but only when the full period is considered, but it is possible these differences are simply random.

\begin{table}[htbp]
\centering
\begin{tabular}{lrr}
\toprule
Portfolio size &  Full period &  GFC  \\
\midrule
10 &       9.16 &       68.6 \\
20 &       7.48 &       70.0  \\
30 &       7.02 &       71.1       \\
\midrule
10 &       8.97 &       69.6  \\
20 &       6.64 &       70.5  \\
30 &       6.65 &       71.0  \\
\bottomrule
\end{tabular}
\caption{Probability (\%) that a random long-short portfolio has a superior Sharpe ratio to a random long-only portfolio for $m=10,20,30$ over the full period and GFC. The top half of the table pertains to uniform sampling, the bottom to stratified sampling. Over the full period, incorporating shorting is unwise, but it carries substantial benefits during a bear market period.}
\label{tab:probcomparison}
\end{table}

\clearpage

\begin{table}[htbp]
\begin{tabular}{lr}
\toprule
Long-only uniformly sampled &         Percentage \\
\midrule
Semiconductors                & 4.67 \\
Oil \& gas                       & 4.43 \\
REIT                          & 4.27 \\
Aerospace defence             & 3.33 \\
Drug manufacturers            & 3.23 \\
\midrule
Auto manufacturers            & 0.50 \\
Leisure                       & 0.50 \\
Retail                        & 0.37 \\
Credit                        & 0.23 \\
Casinos                       & 0.20 \\
\midrule
Short-only uniformly sampled &         Percentage \\
\midrule
Oil \& gas                      & 5.37 \\
Banks - diversified           & 5.30 \\
REIT                          & 4.63 \\
Insurance                     & 4.03 \\
Communications                & 3.30 \\
\midrule
Tobacco                       & 0.60 \\
Internet                      & 0.60 \\
Home improvement retail       & 0.40 \\
Biotechnology                 & 0.30 \\
Internet retail               & 0.20 \\
\midrule
Long-short uniformly sampled &         Percentage \\
\midrule
Semiconductors                & 5.40 \\
Oil \& gas                       & 4.90 \\
REIT                          & 4.53 \\
Banks - diversified           & 3.27 \\
Drug manufacturers            & 3.27 \\
\midrule
Lodging                       & 0.70 \\
Computer hardware             & 0.60 \\
Home improvement retail       & 0.50 \\
Credit                        & 0.47 \\
Internet retail               & 0.27 \\
\bottomrule
\end{tabular}
\caption{Most and least represented sectors by percentage of selection in top performing long-only, short-only and long-short portfolios under uniform sampling ($m=30$). There is a close relationship between the above representation percentages and the corresponding sector sizes.}
\label{tab:unstratified_unadjusted}
\end{table}

\begin{table}[htbp]
\begin{tabular}{lr}
\toprule
Long-only uniformly sampled &         Representation ratio \\
\midrule
Biotechnology                 & 2.02 \\
Internet retail               & 1.97 \\
Non-alcoholic beverages       & 1.94 \\
Restaurants                   & 1.61 \\
Medical distribution          & 1.56 \\
\midrule
Retail                        & 0.49 \\
Department stores             & 0.49 \\
Auto manufacturers            & 0.45 \\
Credit                        & 0.31 \\
Casinos                       & 0.18 \\
\midrule
Short-only uniformly sampled &         Representation ratio \\
\midrule
Retail                        & 2.05 \\
Leisure                       & 1.92 \\
Banks - diversified           & 1.78 \\
Department stores             & 1.68 \\
Casinos                       & 1.55 \\
\midrule
Medical instruments supplies  & 0.58 \\
Waste management              & 0.58 \\
Internet retail               & 0.54 \\
Home improvement retail       & 0.54 \\
Biotechnology                 & 0.27 \\
\midrule
Long-short uniformly sampled &         Representation ratio \\
\midrule
Consumer electronics          & 1.40 \\
Non-alcoholic beverages       & 1.25 \\
Internet                      & 1.25 \\
Restaurants                   & 1.23 \\
Diagnostics research          & 1.23 \\
\midrule
Real estate                   & 0.80 \\
Specialty chemicals           & 0.71 \\
Internet retail               & 0.71 \\
Home improvement retail       & 0.67 \\
Credit                        & 0.63 \\
\bottomrule
\end{tabular}
\caption{Most and least represented sectors by representation of selection in top performing long-only, short-only and long-short portfolios under uniform sampling ($m=30$), adjusted by the sizes of the respective sectors.}
\label{tab:unstratified_adjusted}
\end{table}

\clearpage

\begin{table}[htbp]
\begin{tabular}{lr}
\toprule
Long-only stratified &         Percentage \\
\midrule
Discount stores               & 3.20 \\
Non-alcoholic beverages       & 2.93 \\
Biotechnology                 & 2.77 \\
Medical instruments supplies  & 2.77 \\
Household                     & 2.73 \\
\midrule
Capital markets               & 0.73 \\
Gambling                      & 0.67 \\
Department stores             & 0.60 \\
Retail                        & 0.53 \\
Casinos                       & 0.37 \\
\midrule
Short-only stratified  &         Percentage \\
\midrule
Banks - diversified           & 3.67 \\
Casinos                       & 3.10 \\
Communications                & 2.70 \\
Retail                        & 2.70 \\
Department stores             & 2.70 \\
\midrule
Tobacco                       & 0.90 \\
Diagnostics research          & 0.87 \\
Medical instruments supplies  & 0.80 \\
Internet retail               & 0.73 \\
Biotechnology                 & 0.63 \\
\midrule
Long-short stratified  &         Percentage\\
\midrule
Specialty chemicals           & 2.23 \\
Biotechnology                 & 2.10 \\
Medical instruments supplies  & 2.10 \\
Gaming                        & 1.97 \\
Software infrastructure       & 1.93 \\
\midrule
Advertising                   & 1.37 \\
Credit                        & 1.37 \\
Leisure                       & 1.30 \\
Healthcare plans              & 1.23 \\
Casinos                       & 1.17 \\
\bottomrule
\end{tabular}
\caption{Most and least represented sectors by percentage of selection in top performing long-only, short-only and long-short portfolios under stratified sampling ($m=30$). This aims to adjust for differing sector size in the construction of the sample portfolios.}
\label{tab:stratified}
\end{table}

In Tables \ref{tab:unstratified_unadjusted}, \ref{tab:unstratified_adjusted} and \ref{tab:stratified} we turn to the second part of our experiments, where we investigate the sector composition among top performing portfolios (top 1\% by Sharpe ratio). Table \ref{tab:unstratified_unadjusted} gives the top five most occurring and bottom five least occurring sectors in uniformly sampled long-only, short-only and long-short portfolios of size $m=30$, ranked by the percentage of high-performing portfolios they appear in. In long-only portfolios, we see primarily the largest sectors represented the most: semiconductors, oil \& gas, and REITs (real estate investment trusts).  The smallest sectors (by number of constituent stocks) are represented the least, such as credit and casinos.

Thus, we complement this with Table \ref{tab:unstratified_adjusted}, where we adjust the percentage of representation in top performing portfolios by the size of each sector. For example, biotechnology is represented 2.02 times more often in top performing portfolios relative to the number of biotechnology stocks in our equity universe. Results are expected to be relatively similar to Table \ref{tab:stratified}, which reports percentages of occurrence in top-performing portfolios under stratified sampling. Stratified sampling is specifically designed to correct for the difference in sizes of sectors, so results are expected to be similar. In both Tables \ref{tab:unstratified_adjusted} and \ref{tab:stratified}, the top and bottom sectors, respectively, represent the most and least frequently represented sectors in top portfolios adjusted for their different sizes.

We see two main themes in these aforementioned tables. First, biotechnology, medical distribution and medical instruments are among the most represented sectors (adjusted for size) in top performing long-only portfolios. This indicates the significant growth of the pharmaceutical industry over the past two decades, especially around the COVID-19 pandemic.\cite{statpharma} Secondly, (physical) retail and department stores are among the most represented in short-only portfolios and least represented among long-only portfolios, in stark contrast to internet retail. This reflects the growth in online shopping and the decline of brick-and-mortar retail in the past 20 years. Further, numerous other findings can be observed, such as a poor representation (even after adjusting for sector size) of credit-related companies and casinos, and a strong performance of restaurants and producers of non-alcoholic beverages.

\section{Conclusion}
\label{sec:conclusion}

Our paper makes several contributions to the field. First in Section \ref{sec:Market_structure_shifts}, we introduce new techniques to identify structural shifts in market structure relative to our original sector decomposition, which may identify market dislocations betwen sector pricing. The Wasserstein metric may prove especially useful for online change point detection, given its association with empirically significant financial crises and clarity in the amplitude of its measurements. The four techniques we present each have respective benefits and drawbacks, and could be used by investment managers in different contexts. Based on the individual investor's portfolio management goals and preferences, one of these methods may be preferable to the others. While the metrics $S_t, C_t, W_t$ all reveal financial crises (including signals in advance), the Kendall tau coefficient $K_t$ has a complementary finding, that optimal portfolios must switch between market sectors on a near-continual basis, an observation that is unchanged during crises.

In Section \ref{sec:Network_structure}, we study the market's connectedness and underlying network structure relative to our new sectors, building from both the correlation matrix and transformed distance matrix. We capture two key findings. First, we identify communities of sectors that behave similarly over time. Such communities could be thought of by investment managers as broad themes to diversify across when managing a collection of investments. Their differing market centrality may also be of note and significant when diversifying across the revealed themes. Second, the gold-related sector is identified as an essential group among a portfolio of equity sectors. This is primarily due to gold's long-term (structurally) depressed correlation with almost every sectors, as reflected in the minimum spanning tree of the untransformed correlation matrix.

Finally in Section \ref{sec:Portfolio_sampling}, we conduct an extensive sampling experiment, employing uniform sampling across all stocks and stratified sampling adjusted by sector sizes, and varying portfolio size and the period of analysis. We compare the distributional performance of equally weighted long-only, short-only and long-short investment portfolios by Sharpe ratio, and investigate the sector composition of the top 1\% performing portfolios. We show in quantitative detail that the incorporation of shorting is generally unwise across a long period, but may show positive results over select bear market periods. Regarding sector composition of top performing portfolios, we carefully control for sector size in two different ways, and are able to identify the sectors that were most and least represented in successful portfolios. We see that medical-adjacent industries such as biotechnology, medical instruments and distribution performed well, as did internet retail, while long positions in physical retail were poorly represented in top performing portfolios.

This section is motivated by the study of investment managers along the spectrum of sophistication, which is generally correlated with management and performance fees. Typically, investors who have more discretion as to how they manage their portfolio by way of a wider degree of instruments (equity, credit, options, and others) and direction (both long and short) to use, charge higher fees for managing clients' money. Mathematically, this is equivalent to making investment decisions from a wider universe of potential decisions (a larger decision space). The specific exposures that such sophisticated portfolios can capture may deviate meaningfully from the index in both positive and negative directions, and therefore, make the matter of an investor's skill level extremely important. With the increased fees these portfolio managers command, asset owners should see a commensurate increase in returns and risk-adjusted returns. The general approach in this section could be used by asset allocators to assess fee justifiability among potential fund managers.

Overall, our proposed methodologies are most helpful to a Chief Investment Officer or Head of Quantitative Research type role, where one is managing a portfolio of underlying investors or trading strategies. Our paper demonstrates concrete use cases for managing equity hedge funds, where one can systematically identify shifts in equity market dynamics, clusters of businesses that are likely to respond similarly to exogenous shocks and make appropriate inference as to how portfolio returns are likely to be impacted. These ideas are at the very core of quantitative investment and portfolio management, and although they may seem independent, the three methodologies we present are inter-connected. Our regime identification would be used in an online (real-time) setting to identify changes in market dynamics. As market dynamics change, portfolio managers are likely to change their underlying portfolio composition. To understand which equities are likely to behave similarly and dissimilarly, one must be able to determine their inherent similarity. Our network analysis is a parsimonious framework for identifying clusters of equities that are likely to behave similarly. Finally, our portfolio simulation section provides portfolio managers with an approach for thinking about how the distribution of risk and return is likely to change as the underlying composition changes.

In particular, the methods in this paper, such as the 30-day smoothing window and long-term analyses of portfolio performance, are designed for investors who make portfolio decisions relatively infrequently, that is, investors who are building trading strategies for mid-long term asset allocation. This includes large pensions and endowments, superannuation funds and focused investment groups who specialize in security selection rather than exploiting market inefficiencies in very short term periods. This approach is not uncommon in trading, where momentum and trend following strategies can work based on smoothing out the underlying time series. CTA and managed futures strategies have worked in this way for many years.

There are a variety of opportunities for future research. First, one could extend our work in detecting market structure and see if this can be done in an online setting where machine learning algorithms may predict the probability of future dislocations in the market. Furthermore, many parameters in our modelling approach could be optimized based on the specific properties of an underlying time series. For instance, a time series with more substantial short-term variance may require a tighter window, while a more slow-moving time series may benefit from a larger window. This could be estimated based on the specific time series properties that are presented. Second, it would be interesting to explore the market's inherent connectedness between a mixture of asset classes. That is, how would the market's network diagram look after fixed income, cryptocurrency, currencies, and other options are incorporated into the graph? After including assets with longer-term structurally depressed correlation with equities, alternative assets may be seen as a more suitable diversification strategy than gold. Finally, in our portfolio sampling section, we highlight the required level of manager skill (by quantile of the distribution) in order to justify the existence of long-only vs long-short equity managers. This is due to the enlarged space of outcomes as investors manage portfolios from larger decision spaces. It would be interesting to test this on a wider range of periods than simply the entire window and the GFC. There may be numerous periods other than just market crises where shorting has its place in expanding the possible space of returns in both positive and negative directions. Similarly, with an even larger sample space of asset classes, such as a portfolio that can invest across the capital structure and in different markets, there may be an even higher benchmark on investment manager skill. 


\section*{Data availability}
The data that support the findings of this study are openly available at Ref.  \onlinecite{Yahoo_finance}.

\appendix

\section{List of sectors}
\label{app:sectorslist}

\begin{tabular}{l}
\hline
 Advertising                   \\
 Aerospace defence             \\
 Agriculture                   \\
 Alcoholic beverages           \\
 Asset management              \\
 Auto manufacturers            \\
 Banks - diversified           \\
 Biotechnology                 \\
 Building products \& equipment \\
 Capital markets               \\
 Casinos                       \\
 Communications                \\
 Computer hardware             \\
 Conglomerates                 \\
 Consumer electronics          \\
 Credit                        \\
 Department stores             \\
 Diagnostics research          \\
 Discount stores               \\
 Drug manufacturers            \\
 Electrical equipment          \\
 Entertainment                 \\
 Farming heavy construction    \\
 Footwear \& accessories          \\
 Gambling                      \\
 Gaming                        \\
 Gold                          \\
 Healthcare plans              \\
 Home improvement retail       \\
 Household                     \\
 Industrial metals mining      \\
 Information technologies      \\
 Insurance                     \\
 Internet                      \\
 Internet retail               \\
 Leisure                       \\
 Lodging                       \\
 Medical devices               \\
 Medical distribution          \\
 Medical instruments supplies  \\
 Non-alcoholic beverages       \\
 Oil \& Gas                       \\
 Packaged foods                \\
 Packaging containers          \\
 Publishing                    \\
 REIT                          \\
 Railroads                     \\
 Real estate                   \\
 Restaurants                   \\
 Retail                        \\
 Scientific instruments        \\
 Semiconductors                \\
 Software applications         \\
 Software infrastructure       \\
 Special industrial machinery  \\
 Specialty chemicals           \\
 Telecom                       \\
 Tobacco                       \\
 Utilities                     \\
 Waste management              \\
\hline
\end{tabular}

\section{Wasserstein metric}
\label{app:wasserstein}

In this brief appendix, we explain the definition and computation of the Wasserstein metric $d_W$, which we use between distributions of log returns in Section \ref{sec:Market_structure_shifts}. In greatest generality, the $L^1$ Wasserstein metric is defined between any two probability measures $\mu, \nu$ on a metric space $X$. When $X$ is the set of real numbers $\mathbb{R}$, the metric has the form
\begin{align}
\label{eq:Wasserstein}
    d_W (\mu,\nu) = \inf_{\gamma} \left( \int_{\mathbb{R} \times \mathbb{R}} |x-y| d\gamma  \right).
\end{align}
This infimum is taken over all joint probability measures $\gamma$ on $\mathbb{R}\times \mathbb{R}$ with marginal probability distributions $\mu$ and $\nu$. In the case where $\mu,\nu$ have cumulative distribution functions $F,G$ respectively on $\mathbb{R}$, there is a simple representation:\citep{DelBarrio}
\begin{align}
\label{eq:computeWasserstein}
  d_W (\mu,\nu) =  \left(\int_{0}^1 |F^{-1} - G^{-1}| dx\right),
\end{align}
where $F^{-1}$ is the inverse cumulative distribution function or quantile function associated to $F$. \citep{Gilchrist2000} In our application, we apply $d_W$ to distributions of 30-day log returns $d_W(R_j[t-\tau+1:t]$ and $R_j[t+1:t+\tau])$, both of which yield cumulative distribution functions and quantile functions that are simple to compute.

\bibliography{__NEWNEWREFS}
\end{document}